\documentstyle[preprint,prb,aps]{revtex}
\begin{document}
\draft
\title{Bi-layer Heisenberg model studied by the Schwinger-boson\\
 Gutzwiller-projection method}
\author{T.~Miyazaki, I.~Nakamura, and D.~Yoshioka}
\address{Institute of Physics, College of Arts and
Sciences, University of
Tokyo\\Komaba, Meguro-ku Tokyo 153, Japan\\}
\maketitle
\begin{abstract}
A two-dimensional bi-layer, square lattice Heisenberg model
with different intraplane($J_{\parallel}$) and
interplane($J_{\perp}$) couplings is investigated.
The model is first solved in the Schwinger boson mean-field approximation.
Then the solution is Gutzwiller projected to satisfy the local constraint
that there should be only one boson at each site.
For these wave functions, we perform variational Monte Carlo simulation
up to $24 \times 24 \times 2$ sites.
It is shown that the N\'eel order is destroyed as the interplane coupling
is increased. The obtained critical value, $J_{\perp}/J_{\parallel}=3.51$,
is smaller than that by the mean-field theory.
Excitation spectrum is calculated by a single mode approximation.
It is shown that energy gap develops once the  N\'eel order is destroyed.
\end{abstract}
\section{Introduction}
Spin pseudogap observed in underdoped YBa$_2$Cu$_3$O$_{7-x}$
is one of the fascinating characters of the high-$T_c$ cuprates.
NMR experiments showed that even above the transition temperature of
superconductivity, $T_c$,
static uniform susceptibility and the NMR relaxation rate,$T_1$,
decrease with decreasing temperature.\cite{Takigawa}
Neutron scattering experiments showed the decrease of low energy magnetic
excitation with decreasing temperature and found precursor of a finite
spin gap.\cite{Tranquada}
It has been pointed out that these astonishing experimental results
can be explained
provided that there is a spin pseudogap in the normal state of
high $T_c$ materials.
These phenomena indicating the spin pseudogap, however, have not been
observed in the La$_{2-x}$Sr$_x$CuO$_4$ systems.\cite{LSCO1}
Therefore it is speculated that the number of the CuO$_2$ layers
between the insulating layers is essential for the formation of this gap,
although a successful theory has not been
presented.\cite{Altshuler92,Altshuler94,Ubbens94,Millis3,Sokol}
\par
It is conceivable that the finite concentration of holes affect the spin
configuration and the excitation considerably.
However, as a first step toward understanding of the spin pseudogap behavior,
it is meaningful to study the properties of the bi-layer CuO$_2$ system
at zero doping. Namely we investigate a bi-layer square-lattice Heisenberg
model of spin $1/2$.
\begin{equation}
H = J_{\parallel}  \sum_{i} \sum_{w} \sum_{a=1,2}
    \mbox{\boldmath $S_{i,a} \cdot S_{i+w,a}$}
  + J_{\perp}      \sum_i
    \mbox{\boldmath $S_{i,1} \cdot S_{i,2}$} \quad ,
\end{equation}
where $w=x,y$, $i+w$ represents a site next to the site $i$ in the $w$
direction, and \mbox{\boldmath $S_{i,a}$} is a spin $1/2$ operator at site
$i$ in plane $a$.
The nearest neighbor spins interact antiferromagnetically
with intraplane coupling constant $J_{\parallel}$ and interplane coupling
constant $J_{\perp}$. What we want to know is how the properties of the system
changes as $J_{\perp}/J_{\parallel} \equiv \alpha$ increases:
at what value of $\alpha$ the N\'eel order is destroyed,
and how the excitation spectrum varies.
\par
As for the zero-temperature critical value of $\alpha_c$ for the destruction of
the N\'eel order, there have been several investigations by various methods:
spin wave approximation given by Matsuda and Hida \cite{HidaSW,HidaMSW}
and the Schwinger boson mean-field theory \cite{Millis3} resulted in quite
large critical value, $\alpha_c=4.24$ for the former
and $4.48$ for the latter.
On the other hand, more sophisticated
methods have resulted in much lower critical value.
Quantum Monte Carlo calculation gives it as $2.51 \pm0.01$\cite{QMC} and
the dimer expansions, which is an approach from the
$\alpha \rightarrow \infty$
limit, gives $2.56$.\cite{HidaDim}
One of the aim of the present paper is to obtain this critical value by
another method, the Schwinger-boson Gutzwiller-projection method.
\par
In Schwinger-boson Gutzwiller-projection method we first solve the Hamiltonian
by Schwinger-boson mean-field theory. The obtained ground-state wave function
is Gutzwiller projected to fix the spin at each site to be $1/2$.
The wave function thus obtained is a kind of RVB \cite{RVB1,RVB2}
wave function where
long-range bonds are allowed with amplitude depending on the distance between
the sites.\cite{LDA}
This method was first used by Chen and Xiu for the square lattice
antiferromagnetic Heisenberg model.\cite{Chen}
It was shown that the wave function
obtained this way is quite close to the true ground-state.
This method has also been applied to the anisotropic
Heisenberg model.\cite{MYO}
There it was shown that even in the one-dimensional limit the ground-state
energy, $-0.4377 J$ per site , is quite close to the exact value, $-0.4431 J$
per site.\cite{Bethe}
Therefore we expect that this method gives wave functions quite close
to the actual ground-state in the present system, too. A merit of the present
method is that the wave function is given as an RVB wave function.
Thus vertically coupled dimer state in the limit of
$\alpha \rightarrow \infty$, disordered state in the intermediate value of
$\alpha$,
and the N\'eel state at small $\alpha$ can be described in a unified way by
wave functions with the same structure.
\par
In this paper, using this method we show that the N\'eel order at small
$\alpha$ is destroyed at $\alpha_c=3.51$.
It is expected that gap appears in
the excitation spectrum at $\alpha > \alpha_c$.
This is confirmed by our calculation of the spectrum by a
single mode approximation.
To obtain these results we solve the present
Hamiltonian by the Schwinger-boson mean-field theory in
Sec.\ref{sect:MFT}. The obtained ground-state wave function is
Gutzwiller-projected in Sec.\ref{sect:RVB}.
The single mode approximation for the RVB wave function is discussed
in Sec.\ref{sect:exc}.
In Sec.\ref{sect:results},
we perform variational Monte Carlo simulation for these wave functions
and calculate its energy, spin-spin
correlation, staggered magnetization and low-lying excitation spectrum.
In Sec.\ref{sect:discuss}
critical point and the excitation spectrum are discussed.
\section{Mean-field solution}
\label{sect:MFT}
We introduce four kinds of bose operators,
$s_{i,a,\uparrow}, s_{i,a,\downarrow} (a=1,2)$
to express the spin operators
\begin{equation}
S_{i,a}^{+}=s_{i,a,\uparrow}^{\dagger}s_{i,a,\downarrow} \quad ,
   S_{i,a}^{z}=\frac{1}{2}(s_{i,a,\uparrow}^{\dagger}s_{i,a,\uparrow}
-s_{i,a,\downarrow}^{\dagger} s_{i,a,\downarrow}).
\end{equation}
The commutation relations of the spin operators $\mbox{\boldmath $S_i$}$ are
satisfied in this replacement. We impose a constraint,
\begin{equation}
\label{eqn:2}
s_{i,a,\uparrow}^{\dagger}s_{i,a,\uparrow} +s_{i,a,\downarrow}^{\dagger}
s_{i,a,\downarrow} =1  ,
\end{equation}
in order to guarantee $S=1/2$. Then Hamiltonian is rewritten as follows;
\begin{eqnarray}
\label{eqn:3}
H&=& \frac12 J_{\parallel} \sum_i \sum_w \sum_{a=1,2} \sum_{\sigma}
(s_{i,a,\sigma}^{\dagger}s_{i+w,a,-\sigma}^{\dagger}
s_{i+w,a,\sigma} s_{i,a,-\sigma} -s_{i,a,\sigma}^{\dagger}
s_{i+w,a,-\sigma}^{\dagger} s_{i+w,a,-\sigma} s_{i,a,\sigma})\nonumber\\
&+& \frac12 J_{\perp} \sum_i \sum_{\sigma}
(s_{i,1,\sigma}^{\dagger} s_{i,2,-\sigma}^{\dagger}
s_{i,2,\sigma} s_{i,1,-\sigma} -s_{i,1,\sigma}^{\dagger} s_{i,2,-\sigma}
^{\dagger} s_{i,2,-\sigma} s_{i,1,\sigma}) \nonumber \\
&+&\mu\sum_i \sum_{a=1,2} \sum_{\sigma}s_{i,a,\sigma}^{\dagger}
s_{i,a,\sigma} .
\end{eqnarray}
Here
$\mu$ is a chemical potential introduced to enforce the constraint
Eq.(\ref{eqn:2}) on
the average. To solve the Hamiltonian in the mean-field approximation,
we introduce the following mean-field order parameters
$\Delta _{w,a}$, $\Delta_z$, and $n_{a,\sigma}$
, which give the amplitudes of the intralayer singlet correlations,
interlayer singlet correlations, and an averaged occupation number,
respectively,
\begin{eqnarray}
\label{eq:5}
\Delta_{w} \equiv
\Delta _{w,2}=-\Delta_{w,1}=\frac12 \langle s_{i,2,\downarrow}
  s_{i+w,2,\uparrow} -s_{i,2,\uparrow} s_{i+w,2,\downarrow} \rangle  , \\
\Delta_z=\frac12 \langle s_{i,1,\downarrow} s_{i,2,\uparrow}
              -s_{i,1,\uparrow} s_{i,2,\downarrow} \rangle  , \\
\label{eq:7}
n_{a,\sigma} = \langle s_{i,a,\sigma}^{\dagger} s_{i,a,\sigma} \rangle=
\frac12 .
\end{eqnarray}
After decoupling the Hamiltonian, we rewrite the operator using its Fourier
transformation:
\begin{equation}
s_{i,a,\sigma}=\frac{1}{\sqrt N}\sum_{k} e^{i\mbox{\boldmath $k\cdot r_{i}$}}
s_{k,a,\sigma} \quad ,
\end{equation}
where N is the total number of lattice sites for each layer, and
$\mbox{\boldmath $k$}$ summation is taken over the Brillouin zone
$-\pi \leq k_x \leq \pi,-\pi \leq k_y \leq \pi$. The mean-field Hamiltonian
$H_{\rm MF}$ is written as follows
\begin{eqnarray}
H_{\rm MF} & = & \sum_{k} \sum_{a} \lambda(s_{k,a,\uparrow}^{\dagger}
s_{k,a,\uparrow} +
s_{-k,a,\downarrow}^{\dagger} s_{-k,a,\downarrow})
+ i\gamma_k(s_{k,2,\uparrow}^{\dagger}s_{-k,2,\downarrow}^{\dagger}
 -s_{k,1,\uparrow}^{\dagger} s_{-k,1,\downarrow}^{\dagger})
\nonumber\\
& - & i\gamma_k^{\star}(s_{k,2,\uparrow}
s_{-k,2,\downarrow} - s_{k,1,\uparrow} s_{-k,1,\downarrow})
-i\delta^{\star}(s_{k,1,\uparrow}
 s_{-k,2,\downarrow} - s_{k,2,\uparrow} s_{-k,1,\downarrow})\nonumber\\
& + & i\delta (s_{k,1,\uparrow}^{\dagger}
s_{-k,2,\downarrow}^{\dagger} - s_{k,2,\uparrow}^{\dagger}
s_{-k,1,\downarrow}^{\dagger} )
+ {\rm const.}
 \end{eqnarray}
with
\begin{eqnarray}
\lambda =\mu- J_{\parallel}  ,\\
\gamma_k=2J_{\parallel} (\Delta_x \sin k_x + \Delta_y \sin k_y) ,\\
\delta=-iJ_{\perp}\Delta_z  .
\end{eqnarray}
The
Hamiltonian can be diagonalized by a paraunitary Bogoliubov transformation
\begin{eqnarray}
s_{k,1,\uparrow} = \frac{1}{\sqrt 2}(\cosh\theta_k^+ \alpha_{k\uparrow}-
\cosh \theta_k^-
\beta_{k\uparrow}+i\sinh\theta_k^+ \alpha_{-k\downarrow}^{\dagger}-
i\sinh\theta_k^- \beta_{-k\downarrow}^{\dagger}) ,\\
s_{-k,1,\downarrow}^{\dagger}=\frac{1}{\sqrt 2}(i\sinh\theta_k^+
\alpha_{k\uparrow}
-i\sinh\theta_k^- \beta_{k\uparrow}-\cosh \theta_k^+
\alpha_{-k\downarrow}^{\dagger}
+\cosh \theta_k^- \beta_{-k\downarrow}^{\dagger}) ,\\
s_{k,2,\uparrow}=\frac{1}{\sqrt 2}(\cosh \theta_k^+ \alpha_{k\uparrow}+\cosh
 \theta_k^-
\beta_{k\uparrow}+i\sinh\theta_k^+ \alpha_{-k\downarrow}^{\dagger}
+i\sinh\theta_k^- \beta_{-k\downarrow}^{\dagger}) ,\\
s_{-k,2,\downarrow}^{\dagger}=\frac{1}{\sqrt 2}(-i\sinh\theta_k^+
\alpha_{k\uparrow}
-i\sinh\theta_k^- \beta_{k\uparrow}+\cosh \theta_k^+
\alpha_{-k\downarrow}^{\dagger}
+\cosh \theta_k^- \beta_{-k\downarrow}^{\dagger}) ,
\end{eqnarray}
where
\begin{eqnarray}
\cosh \theta_k^{\pm}&=&\sqrt{\frac{\lambda+E_{k\pm}}{2E_{k\pm}}}\nonumber ,\\
\sinh \theta_k^{\pm}&=&-\sqrt{\frac{\lambda-E_{k\pm}}{2E_{k\pm}}}{\rm sgn}
(\gamma_k \pm \delta),
\end{eqnarray}
\begin{equation}
E_{k\pm}=\sqrt {\lambda^2-(\gamma_k  \pm \delta)^2}  .\\
\end{equation}
After the transformation, Hamiltonian finally becomes
\begin{equation}
H_{\rm MF}=\sum_{k}E_{k+}(\alpha_{k\uparrow}^{\dagger}\alpha_{k\uparrow}+
\alpha_{-k\downarrow}
^{\dagger}\alpha_{-k\downarrow})+E_{k-} (\beta_{k\uparrow}^{\dagger}
\beta_{k\uparrow}+\beta_{-k\downarrow}^{\dagger}\beta_{-k\downarrow})+
{\rm const.}
\end{equation}
The ground-states $|G\rangle $ is defined as the vacuum of the Bose
operator $\alpha_{k\uparrow},\alpha_{-k\downarrow},\beta_{k\uparrow}$, and
$\beta_{-k\downarrow}$, such that
$\alpha_{k\uparrow} | G\rangle=\alpha_{-k\downarrow} | G\rangle=
\beta_{k\uparrow} | G\rangle=
\beta_{-k\downarrow} | G\rangle=0$.

For a finite-size system, the self consistent equations for
$\lambda,\Delta_x,\Delta_y,\Delta_z$ are given
by Eqs.(\ref{eq:5}-\ref{eq:7}),
which lead to
\begin{eqnarray}
\label{eqn:23}
1=\frac{1}{4N}\sum_k \left( \frac{\lambda}{E_{k+}}+\frac{\lambda}{E_{k-}}
\right)  ,\\
\Delta_{w} =\frac{1}{4N} \sum_k \sin k_w
\left(\frac{\gamma_k +\delta}{E_{k+}}+
\frac{\gamma_k-\delta}{E_{k-}}\right)  ,\\
\label{eqn:25}
\Delta_z=\frac{i}{4N} \sum_k \left(\frac{\gamma_k+\delta}{E_{k+}}
                    -\frac{\gamma_k-\delta}{E_{k-}} \right) .
\end{eqnarray}
We find that the free energy takes the same minimal value for
$\Delta_x = \Delta_y$(s-wave)
and $\Delta_x = -\Delta_y$(d-wave).\cite{yd1,yd2}
Since either state gives the same result, we consider only the
s-wave state from now on.
We denote
$\Delta_x=\Delta_y \equiv \Delta_{\parallel}$ and
$-i\Delta_z \equiv \Delta_{\perp}$.
The solution depends on the size of the system $N$. When $N$ is finite,
$E_{k\pm}$ never becomes zero. However, in the limit of $N\rightarrow
\infty$ it is
possible that $E_{k\pm} $ vanishes at
$\mbox{\boldmath $k$}=\mbox{\boldmath $K_{\pm}$}=\pm (\pi/2,\pi/2)$.
In such a case it is known that we need to introduce the Bose condensate
$n_B$, and Eqs.(\ref{eqn:23}-\ref{eqn:25}) are rewritten as
\begin{eqnarray}
1=\frac{1}{4(2\pi)^2} \int_{-\pi}^{\pi}\int_{-\pi}^{\pi}
\left(\frac{\lambda} {E_{k+}}+\frac{\lambda} {E_{k-}} \right)
dk_x dk_y+n_B  ,\\
\Delta_{\parallel}=\frac{1}{4(2\pi)^2} \int_{-\pi}^{\pi}\int_{-\pi}^{\pi}
\sin k_w\left(\frac{\gamma_k+\delta}{E_{k+}}+\frac{\gamma_k-\delta}{E_{k-}}
\right) dk_x dk_y+n_B  ,\\
\Delta_{\perp}=\frac{1}{4(2\pi)^2} \int_{-\pi}^{\pi}\int_{-\pi}^{\pi}
\left(\frac{\gamma_k+\delta}{E_{k+}}
    -\frac{\gamma_k-\delta}{E_{k-}} \right)
dk_x dk_y +n_B.
\end{eqnarray}
When the Bose condensate $n_B$ becomes finite,
we have
$\lambda=4J_{\parallel} \Delta_{\parallel}+J_{\perp}\Delta_{\perp}$.
The self-consistent equations are numerically solved.
Figure~1 shows the $\alpha$ dependence of order-parameters
$\Delta_{\parallel}, \Delta_{\perp}$,
Bose condensate $n_B$, and energy gap $E_g$.
Bose condensate vanishes at $\alpha=4.48$, and
the gap opens for $\alpha>4.48$. The intralayer RVB order parameter,
$\Delta_{\parallel}$, vanishes
at $\alpha=4.62$. For $\alpha>4.62$,
only the interlayer nearest neighbor spin-spin correlation exists.

The intra(inter) layer spin-spin correlation
$\langle\mbox{\boldmath $S_{i,a}\cdot S_{j,a}$}\rangle
(\langle \mbox{\boldmath $S_{i,1}\cdot S_{j,2}$}\rangle)$
 in the grand-state are given as
\begin{eqnarray}
\label{eqn:29}
\langle \mbox{\boldmath $S_{i,a}\cdot S_{j,a}$}\rangle
&=&\frac32 \Bigl[\frac{1}{4N} \sum_k (\frac{\lambda}{E_{k+}}
+\frac{\lambda}{E_{k-}})\cos \mbox{\boldmath $k \cdot r_{i,j}$}+n_B
\cos \mbox{\boldmath $K_+ \cdot r_{i,j}$}\Bigr]^2   \nonumber\\
&&-\frac32 \Bigl[\frac{1}{4N}\sum_k(\frac{\gamma_k+\delta}{E_{k+}}
+\frac{\gamma_k-\delta}{E_{k-}} )
\sin \mbox{\boldmath $k\cdot r_{i,j}$}+
n_B\sin \mbox{\boldmath $K_+\cdot r_{i,j}$}\Bigr]^2  ,
\end{eqnarray}
\begin{eqnarray}
\label{eqn:30}
\langle \mbox{\boldmath $S_{i,1}\cdot S_{j,2}$}\rangle
&=&\frac32 \Bigl[\frac{1}{4N} \sum_k (\frac{\lambda}{E_{k+}}-\frac{\lambda}
{E_{k-}})\sin \mbox{\boldmath $k\cdot r_{i,j}$} +n_B\sin \mbox{\boldmath
$ K_+ \cdot r_{i,j}$}\Bigr]^2\nonumber\\
&&-\frac32
\Bigl[\frac{1}{4N}\sum_k (\frac{\gamma_k +\delta}{E_{k+}}-
\frac{\gamma_k-\delta}
{E_{k-}})\cos \mbox{\boldmath $k\cdot r_{i,j}$} +n_B
\cos \mbox {\boldmath $K_+\cdot r_{i,j}$}\Bigr]^2   ,
\end{eqnarray}
where $\mbox{\boldmath $r_{i,j}=r_{i}-r_{j}$}$.
The summations over $k$ in
Eqs.(\ref{eqn:29},\ref{eqn:30})
vanish in the limit $|r_{i,j}| \rightarrow \infty$.
Therefore,
the correlation extends to infinity only if $n_B >0$, which means
the existence of antiferromagnetic long-range order.
Thus, in the mean-field approximation, the critical point of order-disorder
transition is $4.48$.
\section{RVB wave function}
\label{sect:RVB}
The ground-state wave function obtained in the mean-field theory is expressed
as\begin{eqnarray}
| G\rangle&=&\prod_k \exp \Bigl[ i\frac{\tanh\theta_k^++\tanh\theta_k^-}{2}
(s_{k,2,\uparrow}^{\dagger}
s_{-k,2,\downarrow}^{\dagger}-s_{k,1,\uparrow}^{\dagger}s_{-k,1,\downarrow}^
{\dagger})-\nonumber\\
&&i\frac{\tanh \theta_k^+ -\tanh \theta_k^-}{2}(s_{-k,1,\downarrow}^{\dagger}
s_{k,2,\uparrow}^{\dagger}-s_{k,1,\uparrow}^{\dagger}
s_{-k,2,\downarrow}^{\dagger})\Bigr ] | 0\rangle  ,
\end{eqnarray}
where $|0\rangle$ is the vacuum of the Schwinger bosons.
By the Fourier transformation for $s_{k,1,\uparrow}^{\dagger},
s_{-k,1,\downarrow}^{\dagger}, s_{k,2,\uparrow}^{\dagger},
s_{-k,2,\downarrow}^{\dagger}$,
we can get a real-space representation for this ground-state,
\begin{eqnarray}
| G\rangle=\exp \Bigl[\sum_{i,j}a_{i,j}(s_{i,2,\uparrow}^{\dagger}
s_{j,2,\downarrow}^{\dagger}-s_{i,1,\uparrow}^{\dagger}
s_{j,1,\downarrow}^{\dagger})
+ b_{i,j}(s_{j,1,\downarrow}^{\dagger}
s_{i,2,\uparrow}^{\dagger}-s_{i,1,\uparrow}^{\dagger}
s_{j,2,\downarrow}^{\dagger})
\Bigr] | 0\rangle  ,
\end{eqnarray}
\begin{eqnarray}
\label{eqn:aij}
a_{i,j}=\frac{i}{2N}\sum_k \Bigl[\tanh \theta_k^+ +\tanh \theta_k^-
 \Bigr]
\exp (i\mbox{\boldmath $k\cdot r_{i,j}$}) ,\\
\label{eqn:bij}
b_{i,j}=\frac{-i}{2N}\sum_k
\Bigl[ \tanh \theta_k^+  -\tanh \theta_k^-
\Bigr]
\exp (i\mbox{\boldmath $k\cdot r_{i,j}$})  .
\end{eqnarray}
It is evident that the local constraint, Eq.(\ref{eqn:2})
is not satisfied in this wave function.
We remove this difficulty by projecting the wave function to a space where
each
site is singly occupied. Namely, we perform the Gutzwiller projection, using
 Gutzwiller projection operator $P$,
\begin{equation}
\label{eqn:38}
| G\rangle=P\Bigl[\sum_{i\ne j}a_{i,j}(s_{i,2,\uparrow}^{\dagger}
s_{j,2,\downarrow}^{\dagger}-s_{i,1,\uparrow}^{\dagger}s_{j,1,\downarrow}^
{\dagger})\nonumber \\
+ b_{i,j}(s_{j,1,\downarrow}^{\dagger}
s_{i,2,\uparrow}^{\dagger}
-s_{i,1,\uparrow}^{\dagger}s_{j,2,\downarrow}^{\dagger}
)
\Bigr]^{N}| 0\rangle .
\end{equation}
 From Eq.(\ref{eqn:38})
 it is clear that the ground-state $| G\rangle$ is an RVB state.
The weights of the bond, $a_{i,j}$ and $b_{i,j}$, decay proportional to
$r_{i,j}^{-3}$
except for $b_{i,j}$ at small $J_{\perp}$.
Although it would be possible to regard every $a_{i,j}$ and $b_{i,j}$ as
variational parameters, we here restrict them to be those
given in Eqs.(\ref{eqn:aij},\ref{eqn:bij}).
In the case of $\alpha=0$, this restriction is justified by the result
itself: Chen and Xiu \cite{Chen}
have shown that this choice of $a_{i,j}$ gives excellent
results for the ground-state energy
and the staggered magnetization.
The weights $a_{i,j}$ and $b_{i,j}$ depend on
$\alpha=J_{\perp}/J_{\parallel}$ through the order parameters.
We consider this $\alpha$ in $a_{i,j}$ and $b_{i,j}$ as a variational
parameter.
In order to avoid confusion, we use a symbol $\alpha_{\rm p}$ to mean
the value of $\alpha$ used to obtain the ground-state.

\section{Excitation spectrum}
\label{sect:exc}
Once the approximate ground-state is obtained, excitation spectrum
can be calculated by a method given by Feynman
for liquid $^4{\rm He}$, namely the single mode approximation.\cite{Fyn1,Fyn2}
The essential point of this method is to consider a low-lying
excited state intuitively
and calculate excitation spectrum from a known ground-state.
In our case, low-lying state of this Hamiltonian should be
the spin wave excitation. Thus
we consider the following excited states.
\begin{equation}
| E_{\pm} \rangle = (S_{k,1}^- \pm S_{k,2}^- )| G \rangle \quad ,
\end{equation}
\begin{equation}
S_{k,a}^- \equiv \frac{1}{\sqrt{N} } \sum_{i}
S_{i,a}^- e^{i\mbox{\boldmath $k$} \cdot \mbox{\boldmath $r_{i}$} } \quad ,
\end{equation}
where $| E_{\pm} \rangle$ is variational excited states.
Excitation spectrum, $\omega_{\pm}(\mbox{\boldmath $k$})$, is calculated as
\begin{equation}
\label{eqn:Ek}
\omega_{\pm}(\mbox{\boldmath $k$}) =
\frac{f_{\pm}(\mbox{\boldmath $k$})}
{S_{\pm}(\mbox{\boldmath $k$})} \quad ,
\end{equation}
\begin{equation}
\label{eq:Sk}
S_{\pm}(\mbox{\boldmath $k$})  =
\frac{1}{N} \sum_{i,j}
\langle G | (S_{i,1}^+ \pm S_{i,2}^+ )
            (S_{j,1}^- \pm S_{j,2}^- ) | G \rangle
e^{i\mbox{\boldmath $k$} \cdot \mbox{\boldmath $r_{i,j}$} } \quad ,
\end{equation}
\begin{eqnarray}
\label{eq:fk}
f_{\pm}(\mbox{\boldmath $k$}) & = &
\frac{1}{N} \sum_{i,j}
\langle G | (S_{i,1}^+ \pm S_{i,2}^+ )  [ H, (S_{j,1}^- \pm S_{j,2}^- )]
| G \rangle
e^{i\mbox{\boldmath $k$} \cdot \mbox{\boldmath $r_{i,j}$} } \quad  \nonumber \\
     & = & \frac{J_{\parallel}}{N} \sum_{i,l,\omega^{'}}
           \langle G | (S_{l,1}^+ \pm S_{l,2}^+)
      (-S_{i,1}^- S_{i+\omega^{'},1}^z \mp S_{i,2}^- S_{i+\omega^{'},2}^z
      + S_{i,1}^z S_{i+\omega^{'},1}^-  \pm S_{i,2}^z S_{i+\omega^{'},2}^-)| G
\rangle
     e^{i\mbox{\boldmath $k$} \cdot \mbox{\boldmath $r_{i,l}$} } \nonumber \\
     & + & (1 \mp 1) \frac{J_{\perp}}{N} \sum_{i,l}
      \langle G | (S_{l,1}^+ \pm S_{l,2}^+)
          (S_{i,1}^z S_{i,2}^-  - S_{i,1}^- S_{i,2}^z) | G \rangle
      e^{i\mbox{\boldmath $k$} \cdot \mbox{\boldmath $r_{i,l}$} }   \quad .
\end{eqnarray}
Here,
$\omega^{'}=\pm x, \pm y$, $i+\omega^{'}$ represents a site next to the
site $i$ in the $\omega^{'}$ direction,
$S_{\pm}(\mbox{\boldmath $k$})$ is the static structure factor
and $f_{\pm}(\mbox{\boldmath $k$})$ is a 3-point correlation
function of spin operators. The two modes represent in-phase,
$\omega_{+}(\mbox{\boldmath $k$})$,
and out-of-phase, $\omega_{-}(\mbox{\boldmath $k$})$, spin excitations
of the two layers. \par
%
%
%
%
%
%
Since $|G \rangle$ is an RVB state,
we must consider a loop covering associated with two valence
bond configurations, $|c_1 \rangle, |c_2 \rangle $, to calculate
Eqs.(\ref{eq:Sk},\ref{eq:fk}).\cite{LDA}
For Eq.(\ref{eq:Sk}) we use known results,
\begin{equation}
\frac{ \langle c_1 | S_{i,a}^+ S_{j,b}^- | c_2 \rangle }
{\langle c_1 | c_2 \rangle} =\left \{
  \begin{array} {rl}
  \displaystyle{\frac{1}{2} }& \quad
     \mbox{$(i,a),(j,b)$ belong to the same loop and the same sub-lattice.} \\
  \displaystyle{-\frac{1}{2} }& \quad
     \mbox{$(i,a),(j,b)$ belong to the same loop and different sub-lattice.} \\
  \displaystyle{0} & \quad
     \mbox{$(i,a),(j,b)$ belong to the different loop. }
\end{array} \right.
\end{equation}
For Eq.(\ref{eq:fk}) the following rule is found,
\begin{equation}
\frac{ \langle c_1 | S_{l,a}^+ S_{i,b}^- S_{i+\delta,c}^z | c_2 \rangle }
{\langle c_1 | c_2 \rangle} =
\left \{
  \begin{array} {rl}
  \displaystyle{\frac{1}{4} }& \quad
 \mbox{$(i,b),(i+\delta,c)$ belong to the same loop}\\
  \displaystyle{}& \quad
 \mbox{ and $(l,a)=(i+\delta,c)$.} \\
  \displaystyle{-\frac{1}{4} }& \quad
  \mbox{$(i,b),(i+\delta,c)$ belong to the same loop}\\
  \displaystyle{}& \quad
 \mbox{ and $(l,a)=(i,b)$.} \\
  \displaystyle{0} & \quad
              \mbox{otherwise.}
\end{array} \right.
\end{equation}
Here, $i+\delta$ means the nearest neighbor of $i$-th site.
$S_{\pm}(\mbox{\boldmath $k$})$ can be calculated directly from the first rule.
Using the second rule,
$f_{\pm}(\mbox{\boldmath $k$})$ becomes
\begin{eqnarray}
\label{eq:fk2}
f_{\pm}(\mbox{\boldmath $k$}) & = & \frac{J_\parallel}{N}
(2-\cos{k_x}-\cos{k_y})
\sum_{i} \sum_{w} \sum_{a=1,2}
\langle G | S_{i+w,a}^+ S_{i,a}^- S_{i+w,a}^z | G \rangle \nonumber \\
     & + &  \frac{4(1 \mp 1)J_\perp}{N}
\sum_{i} \langle G | S_{i,2}^+ S_{i,1}^- S_{i,2}^z | G \rangle
 \quad .
\end{eqnarray}
Thus we have only to count the number of the nearest
neighbors in the same loop for each loop covering.
This simplifies the numerical calculation.

\section{Numerical results}
\label{sect:results}
In this section, we show numerical results of the ground-state energy,
spin-spin correlation, staggered magnetization, and the excitation spectrum
as a function of $\alpha$.
We perform Monte Carlo simulations in which RVB states are sampled to
satisfy detailed balance
for lattices with $L \times L \times 2$ sites,
where $L \leq 24$.
All the numerical calculations are performed with periodic boundary
conditions.
For each system size we solve the self-consistent equations
(\ref{eqn:23}-\ref{eqn:25}),
and calculate $a_{i,j},b_{i,j}$ to be used
to construct the wave function at that system size.

\subsection{Ground-state energy}
The energy per site of the bi-layer Heisenberg model,$E$, is given by
the nearest-neighbor spin-spin correlations
$\epsilon_{\parallel}(L,\alpha_p)$
and $\epsilon_{\perp}(L,\alpha_p)$
for a given wave function specified by the parameter
$\alpha_p$:
\begin{equation}
\label{eqn:totE}
  E(L,\alpha_p) = 2J_{\parallel} \epsilon_{\parallel}(L,\alpha_p)
    + \frac{1}{2} J_{\perp} \epsilon_{\perp}(L,\alpha_p) \quad ,
\end{equation}
where
\begin{eqnarray}
  \epsilon_{\parallel}(L,\alpha_p) &=& \frac{1}{4L^2}
  \sum_{i} \sum_w \sum_{a=1,2}
     \langle G | \mbox{\boldmath $S$}_{i,a} \cdot \mbox{\boldmath $S$}_{i+w,a}
                  | G \rangle \quad  , \\
  \epsilon_{\perp}(L,\alpha_p) &=& \frac{1}{L^2}
  \sum_{i}      \langle G |\mbox{\boldmath $S$}_{i,1} \cdot
                            \mbox{\boldmath $S$}_{i,2} | G \rangle \quad .
\end{eqnarray}

To estimate the energy in the thermodynamic limit,
the size dependence is examined
and we find the following size scaling for any fixed $\alpha_{\rm p}$,
\begin{equation}
 \epsilon_{\parallel}(L,\alpha_p) =
          \epsilon_{\parallel}(\alpha_p)
                       + \lambda  L^{-3} + \cdots \quad ,
\end{equation}
\begin{equation}
 \epsilon_{\perp}(L,\alpha_p) =
         \epsilon_{\perp}(\alpha_p)
                       + \lambda  L^{-3} + \cdots \quad ,
\end{equation}
where $\lambda$ is a constant.
This size-scaling coincides with the spin wave theory for a square lattice.
In Fig.~2, $\epsilon_{\parallel}(\alpha_p)$ and
$\epsilon_{\perp}(\alpha_p)$ are shown.
Open circles and solid circles indicate
$\epsilon_{\parallel}(\alpha_p)$ and
$ \epsilon_{\perp}(\alpha_p) $.
Error bars show the standard deviation of the Monte Carlo simulation.
The interplane nearest-neighbor spin-spin correlation $\epsilon_{\parallel}$
has a value of $-0.3333 \pm 0.0006$ at $\alpha_{\rm p}=0$,
which is quite close to the best estimated value of
$-0.3348$.\cite{series,GFMC}
The magnitude of $\epsilon_{\parallel}$ decreases as $\alpha_p$ increases and
finally vanishes at $\alpha_{\rm p}=4.62$.
On the other hand, the magnitude of $\epsilon_{\perp}(\alpha_p)$ increases and
saturates to $0.75$ at $\alpha_{\rm p}=4.62$.
At $\alpha_{\rm p} \geq 4.62$
the intraplane spin correlation vanishes and dimerized state is realized.
\par
The ground-state energy per site at a given $\alpha$ is
calculated as a minimum of
$  E(\alpha_{\rm p}) = 2J_{\parallel} \epsilon_{\parallel}(\alpha_{\rm p})
  + \frac{1}{2} J_{\perp} \epsilon_{\perp}(\alpha_{\rm p})$
with respect to $\alpha_{\rm p}$.
Thus, we can get an optimal variational parameter and energy
for a given $\alpha$.
The relation between the variational parameter($\alpha_{\rm p}$)
and a real coupling($\alpha$) is shown in Fig.~3, and
the ground-state energy per site is shown in Fig.~4.
In Fig.~4 we also show the energy of
dimerized state per site,
$-\frac{3}{8}\alpha J_{\parallel}$,(straight line) for reference.
The difference between the optimal energy and the dimerized energy
becomes smaller with increasing interlayer coupling.


\subsection{Staggered magnetization}
We calculated the spin-spin correlation,
$\langle \mbox{\boldmath $S_{i,a}\cdot S_{j,b}$}\rangle$,
between arbitrary two sites $(i,a)$ and $(j,b)$.
The results
for $ 24 \times 24 \times 2 $ lattice system are shown in Fig.~5,
where absolute value of the intralayer
spin-spin correlation is plotted as a function of
the distance between the two sites.
Open circles are for $\alpha=0.4$
and solid circles are for $\alpha=4.6$. It is obvious that there is a
long-range order at $\alpha=0.4$ and
no long-range order at $\alpha=4.6$.
In the latter case, the correlation decreases
exponentially and the typical correlation length for
the disordered state is of the order of a lattice constant. \par
The long-range order is of the antiferromagnetic one.
In the ordered phase
staggered magnetization of the infinite size system is obtained from
size dependence of the staggered spin-spin correlation
between the mostly separated pairs.
For a given lattice size $L$, we calculated both the intralayer
correlation $M_0(L)^2$,
and interlayer correlation $M_1(L)^2$: 
\begin{eqnarray}
   \label{eqn:cl2l2}
M_0(L)^2    & \equiv &  \frac{1}{2N} {\sum_{i,j}}^{'} \sum_{a=1,2}
     \langle | \mbox{\boldmath $S$}_{i,a} \cdot \mbox{\boldmath $S$}_{j,a}  |
                 \rangle  \quad , \\
M_1(L)^2    & \equiv &  \frac{1}{N} {\sum_{i,j}}^{'}
  \langle | \mbox{\boldmath $S$}_{i,1} \cdot \mbox{\boldmath $S$}_{j,2}  |
                 \rangle  \quad ,
\end{eqnarray}
where the summation is taken for all the pair of $i$ and $j$ such that
$\mbox{\boldmath $r_{i}-r_{j}$} = (\pm L/2, \pm L/2)$.

Except at $\alpha=0$, $M_0(L)$ and $M_1(L)$ coincide within the Monte Carlo
statistical error.
As shown in Fig.~6, they are well fitted by the size scaling,
\begin{equation}
\label{eqn:mzl}
M_0(L) = M_1(L)  = M(\infty) +  \mu L^{-1} + \cdots \quad ,
\end{equation}
where $\mu$ is a constant.
This scaling agrees with the prediction of the spin wave theory and
arguments given by Huse.\cite{Huse}
The staggered magnetization $M_0=M(\infty)$ as a function of $\alpha$
is given in Fig.~7. In this figure,
the results of the mean-field theory(MFT) are also shown.
In the case of small $\alpha$, the interlayer coupling enhances the
antiferromagnetic long-range order. This is because the system acquires
a weak three-dimensionality and quantum fluctuation is suppressed.
On the other hand, for larger $\alpha$,
the magnetizations are suppressed. This behavior is consistent with the
result of Matsuda and Hida in the spin-wave theory.\cite{HidaSW}
The staggered magnetization vanishes at $\alpha_c=3.51 \pm 0.05$.

\subsection{Excitation spectrum}
We calculate the structure factor, $S_{\pm}(\mbox{\boldmath $k$})$,
and excitation spectrum, $\omega_{\pm}(\mbox{\boldmath $k$})$
as a function of coupling
$\alpha$. The calculations are done for $ 24 \times 24 \times 2 $ lattice.
The behavior of $S_{\pm}(\mbox{\boldmath $k$})$ and
$\omega_{\pm}(\mbox{\boldmath $k$})$
strongly depends
on whether the system has
long-range order or not.
The result of $S_{\pm}(\mbox{\boldmath $k$})$ is shown in Fig.~8 and
$\omega_{\pm}(\mbox{\boldmath $k$})$ is shown in Fig.~9.
Here, three typical couplings are taken; $\alpha=0.4$(open circles),
$\alpha=2.4$(closed circles), $\alpha=3.6$(open squares).
The third coupling is for the system in the disordered phase.
For each figure, (a) is for the plus mode and (b) is for the minus mode, and
$\Gamma=(0,0)$, $X=(0,\pi)$, $M=(\pi,\pi)$ in momentum space.
\par
It is obvious from Fig.~8 that $S_{+}(\mbox{\boldmath $k$})$ of
the ordered state($\alpha=0.4,2.4$) is proportional to $k$ near
$\Gamma$ point and
$S_{-}(\mbox{\boldmath $k$})$ has an antiferromagnetic peak
at $M$ point.
On the other hand, $S_{+}(\mbox{\boldmath $k$})$
at $\alpha=3.6$ increases quadratically with
$k$ near $\Gamma$ point.(See inset of Fig.~8(a)).
In the N\'eel state, the excitation is gapless at two points.
One is $\omega_{+}(\mbox{\boldmath $k$})$ at $\Gamma$ point.
Around this point,
since the function $f_{+}(\mbox{\boldmath $k$})$ in Eq.(\ref{eqn:Ek})
behaves like $f_{+}(\mbox{\boldmath $k$}) \propto k^2$ and the structure
factor does $S_{+}(\mbox{\boldmath $k$}) \propto k$,
the excitation is proportional to $k$.
The other is $\omega_{-}(\mbox{\boldmath $k$})$
at $M$ point where $S_{-}(\mbox{\boldmath $k$})$ diverges
due to the antiferromagnetic long-range order.
Thus, the gap opens when the structure factor becomes proportional to
the square of $k$ for the former point and when the structure factor does
not diverge, that is, the system becomes the disordered state
for the latter point.
In the former case, we should determine the critical coupling, $\alpha_{c2}$,
where the gap opens.
We take five $k_x$ points and do the following
fitting along the $\Gamma-X$ line;
$S_{+}(k_x)=a_1 k_x + a_2 k_x^2 + a_3 k_x^3 $,
where $a_1,a_2$,and $a_3$ are fitting parameters.
The result of $a_1$ versus $\alpha$ is shown in Fig.~7
(open squares).
Comparing the result of staggered magnetization with this coefficient,
we find that the critical point $\alpha_{c2}$ is equal to the $\alpha_c$
within the statistical and fitting errors, which gives
$\alpha_c=3.51 \pm 0.05$.
The $\alpha$ dependence of the gap is shown in Fig.~10.
All values are scaled by $J_{\parallel}$. Open circles are for
$\omega_{+}(0,0)$ and closed circles are for
$\omega_{-}(\pi,\pi)$.
In the disordered phase,
excitation energy $\omega_{-}(\pi,\pi)$ always takes
smaller value.
Spin wave velocity along the $\Gamma-X$ line is calculated
for the ordered state
and the result is shown in the inset of Fig.~10.
Here, $Z_c$ is the renormalization factor. Namely spin wave velocity
is given by $\sqrt{2} Z_c J_{\parallel}$.
As the coupling increases, the velocity first
slightly decreases and then suddenly increases near the critical point.

\section{DISCUSSIONS}
\label{sect:discuss}

In this paper we first solved the Hamiltonian by the Schwinger-boson
mean-field theory.
Then the solution is Gutzwiller projected to obtain variational ground state
wave functions, which are examined by the Monte Carlo
simulation for finite sizes.

We first see the advantage of our variational Monte Carlo simulation.
In the mean-field calculation, the system becomes dimerized for
$\alpha > 4.62$. For this region, interlayer order parameter,
$\Delta_{\parallel}$, is zero and only dimer coupling between the
layers is permitted. In addition, excitation spectrum becomes flat in
momentum space;
$E_g(k) = \sqrt{\lambda^2 - \delta^2 }$.
As a matter of fact, theoretically, this state
must be only realized at $\alpha \rightarrow \infty$.
This disadvantage is removed in Monte Carlo simulation.
It is estimated from Fig.~4 that the virtual critical point
where the system stabilizes with the dimerized state is $11.0$.
This means that the Gutzwiller projection improves the wave functions.
The improved wave function can describe the disordered
state without dimerization at least up to $\alpha=11.0$. \par
There have been many investigations for the order-disorder critical point.
Our mean-field result is essentially same as the previous report\cite{Millis3}
and the modified spin wave theory.\cite{HidaMSW}
These give the critical value of $\alpha$ around $4.5$.\cite{foot2}
This value is much larger than the results by the other methods:
$2.56$ by the dimer expansion, and $2.51 \pm 0.01$ by the quantum Monte
Carlo method. However, these latter values are still formidably larger than
the value of $\alpha$ realized in the bi-layer cuprates.
Our motivation for this work was to see if our method gives the critical
value closer to the experimental value or not. Our result,
$\alpha_{c}=3.51 \pm 0.05$,\cite{Note} is not for this expectation, and
confirms the previous theories that without doping bi-layer Heisenberg model
will not give an explanation for the spin gap behavior of the experiments.
\par
We also calculated the excitation spectrum,
especially for the disordered phase.
It is not obvious whether the system has always a finite gap in disordered
state. For instance, there is no long-range order for the one-dimensional
$s=1/2$ antiferromagnetic Heisenberg model, though the excitation
spectrum is gapless.
In our bi-layer two-dimensional Heisenberg model,
we find that there is always a finite gap for the disordered state.
Within the statistical and fitting errors,
it occurs at $\alpha_c=3.51 \pm 0.05$.
In disordered region, the spin-spin correlation decays as the distance
exponentially. The structure factor,
$S_{-}(\mbox{\boldmath $k$})$, near the critical coupling,
however, has a large maximal value at $M$ point,
which makes the excitation spectrum be minimized
at that point. This shows that even in the disordered state
the antiferromagnetic spin fluctuation is strong.
It should be remarked that even though the spectrum
$\omega_{-}(\mbox{\boldmath $k$})$ at $\alpha=3.6$
looks singular at $\mbox{\boldmath $Q$}=(\pi,\pi)$, this is not the case.
Around $\mbox{\boldmath $Q$}$ it should be quadratic in
($\mbox{\boldmath $k-Q$}$). Such a behavior is not apparent in Fig.~9(b)
due to
the lack of data close enough to $\mbox{\boldmath $Q$}$.
\par
At $\alpha=0$ where
the model becomes the single layer Heisenberg model,
our result can be compared with other methods:
spin wave theory, series expansions and single mode
approximation.\cite{Igarashi,SEbi,SMAbi}
Our result of the spectrum is roughly proportional to those of other methods
over the entire Brillouin zone.
The maximal value is around 2.65$J_{\parallel}$
at $X$ or $L=(\pi/2,\pi/2)$ point.
Series expansions predicted the maximum is about 2.35$J_{\parallel}$
at $L$ point \cite{SEbi}
and
single mode approximation based on the expansions around the Ising limit
estimated the maximum about $2.5J_{\parallel}$ at $L$ point.\cite{SMAbi}
Both results are close to our result.
The most remarkable difference from the other methods
is the spin wave velocity.
The renormalization factor, $Z_c$, is $1.99 \pm 0.03$ at $\alpha=0$
which is $1.69$ times larger than the best estimated value around
$1.18 \pm 0.02$.\cite{review}
This difference indicates that multi-magnon contribution to
$S_{+}(\mbox{\boldmath $k$})$ is not
negligible. However, since this method gives qualitatively correct behavior,
we believe it gives qualitatively correct spectrum at $\alpha >0$ also.
Finally we remark that the non-monotonous behavior of spin wave velocity
with increasing interlayer coupling can be
understood from that of the coefficient($a_1$) of the structure factor
shown in  Fig.~7, since the spin wave velocity is inversely proportional to
$a_1$.

In conclusion, we have investigated the bi-layer Heisenberg model using the
Schwinger-boson Gutzwiller-projection method.
We find that there is an order-disorder transition with increasing
interlayer coupling.
The critical point is $\alpha_c = 3.51 \pm 0.05$.
Excitation spectrum can be calculated for wide range of
coupling and we find that
the spin excitation has always a finite gap for disordered phase and
the minimum of the spectrum is located at $M$ point.
Our model corresponds to the half-filled case for high-$T_c$
cuprates.
Although $\alpha_c$ in this case is quite large, it is possible that hole
doping reduces the value extremely.
Then it will be possible that our disordered state continuously changes
into the spin gap state.
The similar treatment for a hole doped model, $t-t'-J$ model,
is our next problem.


\section*{Acknowledgements}
\vspace{0.2cm}
The authors thank M.~Ogata for useful comments on the results of our
Monte Carlo simulations.


%

\newpage
\section*{FIGURE CAPTIONS}
FIG.~1. \quad Mean-field values of order parameters $\Delta_{\parallel},
\Delta_{\perp}$,
Bose-condensate $n_B$ and energy gap $E_g$ as a function of $\alpha$. \par
\vspace{1cm}
FIG.~2. \quad The nearest neighbor spin correlation for each direction is
shown.
Open circles are for $ \epsilon_{\parallel} $, and
solid circles are for $  \epsilon_{\perp} $.
Error bars result from Monte Carlo statistical errors. \par
\vspace{1cm}
FIG.~3. \quad
The variational parameter $\alpha_{\rm p}$ which minimizes the ground
state energy for a given parameter $\alpha$.\par
\vspace{1cm}
FIG.~4. \quad  Total energy per site as a function of $\alpha$.
Open circles are for variational Monte Carlo results and
straight line is for dimerized state, $-0.375\alpha$.
\par
\vspace{1cm}
FIG.~5.  \quad  Spin-spin correlation for $\alpha=0.4$(open circles) and
$\alpha=4.6$(solid circles).
Each calculation is done for $24 \times 24 \times 2 $ lattice. Here,
$r_{i,j}$ means the distance between two sites.
It is obvious that there is a long-range order
for $\alpha=0.4$ but no long-range order for $\alpha=3.6$.
\par
\vspace{1cm}
FIG.~6. \quad  $M_0(L)$ versus $1/L$  for
$\alpha=0.0,0.8,1.7,3.1$, and $4.6$.
\par
\vspace{1cm}
FIG.~7.  \quad  Staggered magnetization as a function of $\alpha$.
Open circles are for the mean-field theory, and
solid circles are for variational
Monte Carlo results.
The magnitude of the $k$-linear term in the expansion of
$S_{+}(\mbox{\boldmath$k$})$ around the $\Gamma$ point, $a_1$, is
also shown by open squares.
\par
\vspace{1cm}
FIG.~8. \quad  The structure factors (a)$S_{+}(\mbox{\boldmath$k$})$
and (b)$S_{-}(\mbox{\boldmath$k$})$.
Open circles, closed circles, and open squares are for
$\alpha=0.4$, $2.4$, and $3.6$, respectively.
Inset shows the detailed structure of $\alpha=3.6$ along
the $\Gamma-X$ line.
Note that the value at $M$ point of $S_{-}(\mbox{\boldmath$k$})$ is
too large to be shown in the figure.
\par
\vspace{1cm}
FIG.~9. \quad
Excitation spectrum (a)$\omega_{+}(\mbox{\boldmath$k$})$
and (b)$\omega_{-}(\mbox{\boldmath$k$})$.
The same values for $\alpha$ are chosen and indicated by the same symbols as
in Fig.~8.
At $\alpha=3.6$, gap opens at $\Gamma$ point for
$\omega_{+}(\mbox{\boldmath$k$})$
and at $M$ point for $\omega_{-}(\mbox{\boldmath$k$})$.
\par
\vspace{1cm}
FIG.~10. \quad
The $\alpha$ dependence of the gap for $\omega_{+}(0,0)$
and $\omega_{-}(\pi,\pi)$.
In the inset, the renormalization factor of the linear spin wave
velocity, $Z_c$, is also shown.


\begin{thebibliography}{99}
\bibitem{foot2}Schwinger boson mean-field theory and the modified spin wave
theory are essentially the same. In the present system, the latter has given
the smaller critical value
because the possibility of the first order phase transition
is considered.
\end{thebibliography}
\end{document}